\documentclass[twocolumn,aps,prc,superscriptaddress,showpacs]{revtex4}
\usepackage{amsmath,bm}%
\usepackage{graphicx}%

\begin{document}

\draft
\title{ $\Delta$-scaling and Information Entropy in Ultra-Relativistic                     
Nucleus-Nucleus Collisions 
}
\thanks{ Supported by the Major State Basic Research Development Program under Contract No G200077400, the 
National Natural Science Foundation of China under Grant No 10135030, and the National Natural Science 
Foundation of China for Distinguished Young Scholars under Grant No 19725521}

\author{ G. L. Ma}
\affiliation{Shanghai Institute of Nuclear Research, Chinese Academy of Scienecs, P.O. Box 800-204,
Shanghai 201800, China}
\author{ Y. G. Ma}
\thanks{Corresponding author. Email: ygma@sinr.ac.cn}
\affiliation{Shanghai Institute of Nuclear Research, Chinese Academy of Scienecs, P.O. Box 800-204,
Shanghai 201800, China}
\author{K. Wang}
\affiliation{Shanghai Institute of Nuclear Research, Chinese Academy of Scienecs, P.O. Box 800-204,
Shanghai 201800, China}
\author{B. H. Sa}
\affiliation{ China Institute of Atomic Energy, P.O.Box 918, Beijing 102413, China}
\author{W. Q. Shen}
\affiliation{Shanghai Institute of Nuclear Research, Chinese Academy of Scienecs, P.O. Box 800-204,
Shanghai 201800, China}
\author{H. Z. Huang}£¬
\affiliation{ Dept of Physics, University of California at Los Angeles, CA 90095, USA}
\author{X. Z. Cai}
\author{H. Y. Zhang}
\author{Z. H.  Lu}
\author{C. Zhong}
\author{J. G.  Chen}
\author{Y. B.  Wei}
\author{X. F.  Zhou}

\affiliation{Shanghai Institute of Nuclear Research, Chinese Academy of Scienecs, P.O. Box 800-204,
Shanghai 201800, China}

\date{\today}

\begin{abstract}
The $\Delta$-scaling method has been applied to ultra-relativistic p+p, C+C and Pb+Pb 
collision data simulated using a high energy Monte Carlo package, LUCIAE 3.0. The 
$\Delta$-scaling is found to be valid for some physical variables, such as charged particle 
multiplicity, strange particle multiplicity and number of binary nucleon-nucleon 
collisions from these simulated nucleus-nucleus collisions over an extended energy 
ranging from $E_{lab}$ = 20 to 200 A GeV. In addition we derived information entropy 
from the multiplicity distribution as a function of beam energy for these collisions. 
\end{abstract}


\maketitle


Ultra-relativistic heavy ion collisions provide a unique means to search for a new 
state of matter, the Quark-Gluon Plasma (QGP), where quarks and gluons over an 
extended volume are de-confined [1-3] and a phase transition between hadrons and 
the QGP will occur [4-6]. There could be a discontinuity in global features, 
in particular, the entropy from particle multiplicities, in nuclear collisions associated 
with the onset of the phase transition. In this Letter we report the first application of 
the $\Delta$-scaling method and information entropy calculation in ultra-relativistic heavy 
ion collisions. 

In intermediate energy heavy ion collisions,  $\Delta$-scaling was proposed by Botet 
and Ploszajczak [7]. They applied the $\Delta$-scaling law to the INDRA data in intermediate 
energy heavy ion collisions (Xe+Sn, 25-100 MeV/nucleon) by using $Z_{max}$ (the 
maximum of charge in reactions) as order parameter [8] and found that the 
distributions of $Z_{max}$ obey  the $\Delta$ = 1/2 scaling law below 32 MeV/nucleon while they 
obey the $\Delta$ = 1 scaling law above 32 MeV/nucleon. This indicates that a transition from 
an order phase to the maximum fluctuation phase (disorder phase) occurs around 32
MeV/nucleon. Recently, Ma et al. analysed the multi-fragmentation data for mass around  
$\sim$ 36 light nuclei systems. They found a similar change of  $\Delta$-scaling occurs when 
the excitation energy of the system around 5.6 MeV/nucleon. Combination of analysis with 
other probes of phase transition (eg., Zipf law [9] and maximum fluctuations [10]) 
reflects a transition of matter phases with the change of  $\Delta$-scaling [11].  
Since $\Delta$-scaling is a useful tool to identify the phase transition, 
we try to make the similar analysis in ultra-relativistic heavy-ion collisions. 

Botet and Ploszajczak proposed $\Delta$-scaling to identify the transition in  
intermediate energy heavy ion collisions [12,13].  The $\Delta$-scaling law is observed when 
two or more probability distributions $P[m]$ of the stochastic observable $m$ collapse 
onto a single scaling curve  $\Phi$(z) if a new scaling observable is defined by:

\begin{equation}
    z  = \frac{(m-m^*)}{\langle m\rangle^\Delta}
\end{equation}

This curve is:
\begin{equation}
\langle m\rangle^\Delta P[m] = \Phi (z) = \Phi [\frac{m-m^*}{\langle m\rangle^\Delta}]                 
\end{equation}
where  $\Delta$ is a scaling parameter, $m^*$ is the most probable value of $m$, and $\langle m\rangle$ is the 
mean of $m$. When $\Delta$ = 1, this kind of scaling law is called the first scaling law that is 
caused by self-similarity of system. This self-similarity means that, if these distributions 
with different $\langle m\rangle$ by a new kind of variable $z$, they entirely collapse on the same 
curve. In fact, the famous KNO scaling [14] is the special case that the  $\Delta$=1 scaling 
law holds with a stochastic observable of multiplicity of particles. The INDRA data was 
explored using  the $\Delta$ = 1/2 and 1 scaling laws and a phase transition was observed [15]. If 
we assume that $P[m]$ is a Gaussian distribution, we have:
\begin{equation}
         P [m]  =  \frac{1}{\sigma \sqrt{2\pi}} exp[-\frac{1}{2} (\frac{m-\mu}{\sigma})^2  ]                             
\end{equation}
where $\mu = \langle m\rangle = m^*$ , $\sigma$ is the width of the Gaussian distribution, they both depend on 
incident energy. If this Gaussian distribution $P[m]$ obeys $\Delta$-scaling law , we 
should have:
\begin{equation}
      \mu^\Delta \propto \sigma                                          
\end{equation}

On the other hand, the information entropy is  observable such that it was proposed to 
describe the fluctuation and disorder of a system by Shannon [16]. It is defined by:
\begin{equation}
          H  = -\int P(m) lnP(m) dm                                
\end{equation}      
where $\int P(m) dm$ = 1. This indicates that the quantity of fluctuation of system and depend  
on the distribution $P[m]$. If $P[m]$ is the Gaussian distribution, we have :
\begin{equation}
H  =  ln\sigma +(1+ln2\pi)/2 \simeq ln\sigma + 1.419                
\end{equation}

In this work, we investigate the $\Delta$-scaling law and the information entropy for p+p, C+C 
and Pb+Pb in high-energy collisions (20-200 AGeV) with help of LUCIAE 3.0 of Sa and 
Tai [17]. The head-on collisions are simulated in this work.  The LUCIAE is of a Monte 
Carlo model and is an extension of the FRITIOF [18]. Here the nucleus - nucleus collision 
is described as the sum of nucleon-nucleon collisions. LUCIAE was improved 
in the following three aspects: (1) Re-scattering of final hadrons, spectator and participant 
nucleons are considered in the LUCIAE, because the role of re-scattering [19] can not be 
neglected in high-energy domain. (2) The LUCIAE implements the Firecracker Model that 
includes collective multi-gluon emission from the color fields of interacting strings in 
the early stage of relativistic ion collisions. (3) LUCIAE introduces the suppression of 
strange quarks and effective string tensor to make some parameters related to product 
of strange quarks and string tensor in the JETSET depend on incident energy, size of 
system, centrality etc. Generally the LUCIAE can explain many data very well [20,21].

First, we choose the particle multiplicity produced in each event as stochastic 
observable $m$ and simulated 10000 events for p+p at every energy point to obtain 
normalized multiplicity distributions correspondingly. Then we performed  $\Delta$-scaling.  
As a result, we find that the multiplicity distributions are approximately satisfied 
with the $\Delta$ = 1 scaling law, as shown in Figure 1.
   
\begin{figure}
\includegraphics[scale=0.35]{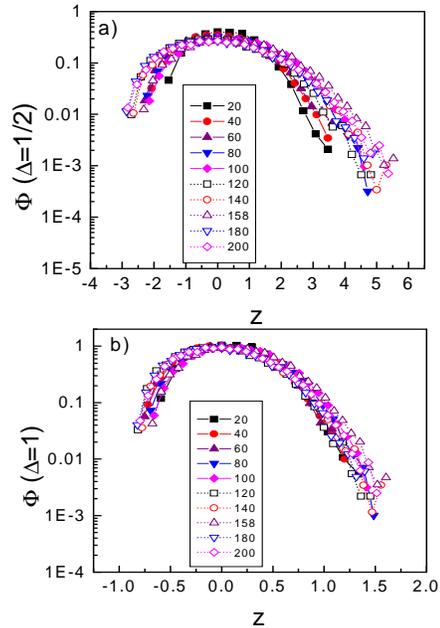}
\caption{\footnotesize $\Delta$ = 1/2 (a) and  $\Delta$ = 1 (b) scalings 
for the multiplicity distribution of p + p
reactions for  the different beam energies.}
\label{Cv_pi}
\end{figure}

We calculated  $\langle m\rangle = \sum m_i P(m_i)$ and $m_{RMS} = \sqrt{\sum (m_i^2 -(m^*)^2) P(m_i)}$ in 
terms of  these normalized multiplicity distributions directly. While we fitted these 
distributions with Gaussian distributions to obtain $\mu$ and $\sigma$. By comparing $\mu$ 
with $\langle m\rangle$ and $\sigma$ with $m_{RMS}$, we found that these distributions are basically Gaussian. 
Table 1 shows the fitting parameters and information entropy ($H_{direct}$ or $H_{gauss}$, which  
was calculated directly or by Gaussian parameter, respectively (see details in Ref. [22] and 
in the following).

\begin{table}
  \begin{tabular}{lllllll}
  \toprule
   E(AGeV) & $H_{direct}$ & $H_{gauss}$ & $\langle m\rangle$ & $\mu$ & $\sigma$ & $m_{RMS}$\\
  \colrule

   20 & 2.237 & 2.290&7.374&7.332&2.390&2.330\\
   \colrule
   40&2.512&2.553&8.820&8.724&3.107&3.032\\
  \colrule
   60&2.634&2.687&9.767&9.638&3.555&3.470\\
  \colrule
   80&2.747&2.797&10.76&10.60&3.968&3.880\\
  \colrule
   100&2.791&2.855&11.24&11.11&4.202&4.127\\
   \colrule
  120&2.841&2.893&11.65&11.54&4.365&4.320\\
    \colrule
 140&2.912&2.921&12.18&12.25&4.489&4.617\\
   \colrule
  160&2.935&2.925&12.56&12.54&4.508&4.633\\
    \colrule
 180&2.965&2.986&12.55&12.48&4.792&4.783\\
  \colrule
   200&3.008&3.030&13.11&12.92&5.010&4.951\\
\botrule
\end{tabular}
\caption{Parameters in the direct calculation and fits for p + p.}
\label{tab1}
\end{table}

In order to better investigate the $\Delta$-scaling, we defined a coefficient 
\begin{equation}
            L  =  \frac{\langle m\rangle^\Delta}{ m_{RMS}} , 
\end{equation}                           
which characterizes the validity of the $\Delta$-scaling. We investigat its dependence of 
incident energy with different $\Delta$ values. Figure 2 shows that the $L$-dependence on 
incident energy from  $\Delta$ = 0.5 to  $\Delta$ = 1.6 . As a result, we find that $L$ 
is nearly a constant in whole investigated range (20-200 AGeV) when  $\Delta$=1.35, 
i.e., the system obeys the $\Delta$-scaling law most suitably with  $\Delta$ = 1.35 (see Fig.3a).

\begin{figure}
\includegraphics[scale=0.35]{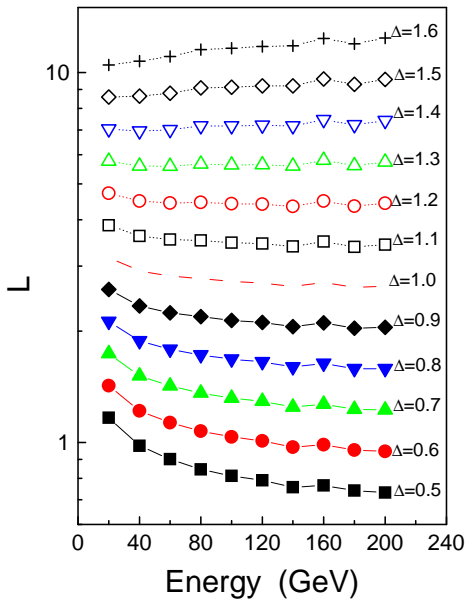}
\caption{\footnotesize $L$ as a function of energies.}
\label{fig2}
\end{figure}

We also deal with another two systems of C+C (5000 events) and Pb+Pb (1000 
events). Similarly, their multiplicity distributions basically obey Gaussian 
distributions. In the same way, the best  $\Delta$-scaling is obtained with $\Delta$ = 1.00 for C+C 
and $\Delta$ = 0.80 for Pb + Pb, respectively. Figure 3b and 3c illustrate these results.

\begin{figure}
\includegraphics[scale=0.35]{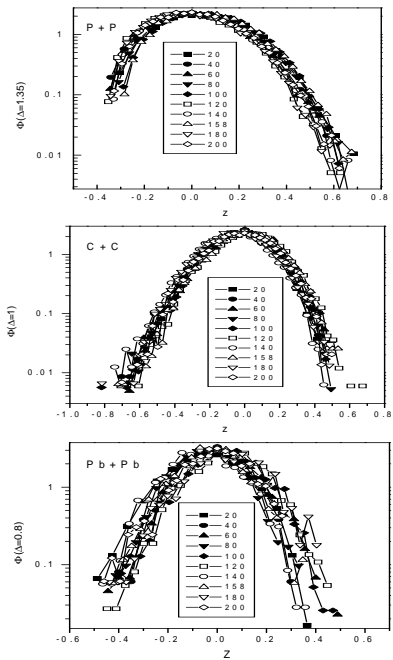}
\caption{\footnotesize $\Delta$-scaling for p + p with  $\Delta$ = 1.35 (upper panel), 
for  C + C with  $\Delta$ = 1.00 (middle pannel) and for  
Pb+Pb with  $\Delta$ = 0.80 (lower pannel) for the different beam energies.}
\label{fig3}
\end{figure}

Through  $\Delta$-scaling for different systems, we found that the distributions of  
particle multiplicity obey Gaussian distributions approximately, especially for C + C 
and Pb + Pb systems and they obey the $\Delta$-scaling law in a wide energy range for a given 
system, which indicates that no phase transition and no change of reaction mechanism exist. 
This is expected for the simulated data because the underlying particle production 
dynamics in the LUCIAE is a smooth function of beam energy without a phase transition.

We calculate the respective information entropy in terms of these distributions of 
particle multiplicity with the method proposed by Ma in Ref. [22]. Figure 4 shows the 
dependences of the information entropy on incident energy for the p + p, C + C and Pb + Pb 
systems. The information entropy increases with the incident energy and with the sizes of the 
system monotonously. Also, the calculated values of $H_{direct}$ are consistent with the value 
$H_{gauss}$ obtained from Eq. (6) in the Gaussian distribution. Again, 
 no indication of phase transition exists.

\begin{figure}
\includegraphics[scale=0.35]{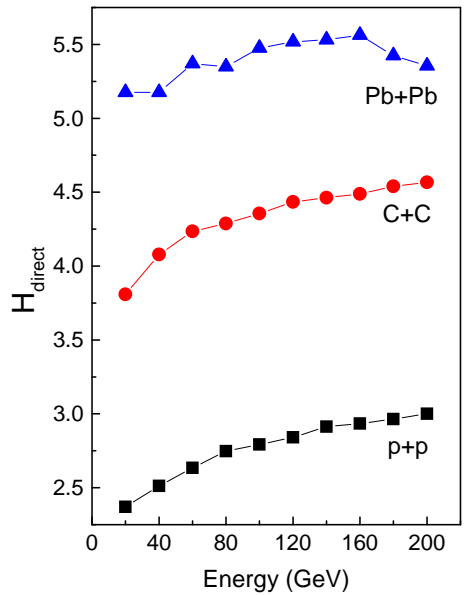}
\caption{\footnotesize Energy dependence of information entropy for p + p, C + C 
and Pb + Pb, respectively.}
\label{fig4}
\end{figure}

In addition, we use the distribution of strange particles multiplicity [23] and the 
distribution of the number of binary collisions [24] as two stochastic observables to 
make  the $\Delta$-scaling plots and extract the information entropy in the same way. The similar 
 $\Delta$-scaling was obtained and a monotonous information entropy was also observed. 

In summary, we have demonstrated, for the first time to our knowledge, the $\Delta$-scaling of charged 
particle multiplicity, strange particle multiplicity and the number of binary collisions 
using simulated p + p, C + C and Pb + Pb collisions from $E_{lab}$ = 20 to $E_{lab}$ = 200 AGeV. The 
LUCIAE 3.0, which includes the final state interactions, was used for the simulation. 
The $\Delta$-scaling means that these observables obey a certain kind of universal laws,  
regardless of beam energy and collision system. We found that the scaling values 
$\Delta$ for charged particle multiplicity distributions are 1.35, 1.00 and 0.80 for 
p + p, C + C and Pb + Pb collisions, respectively. Moreover, the information entropy calculated 
from charged multiplicity distributions increases with the beam energy and with the 
colliding system size monotonously. Both the $\Delta$-scaling and the entropy values show 
no dis-continuity as a function of beam energy as expected because the LUCIAE has no 
change of particle production dynamics, while they are associated with a phase transition in 
the simulated data. We expect that the $\Delta$-scaling and the entropy variable can be a 
valuable tool to search for possible dis-continuities in nucleus-nucleus collisions 
associated with the onset of a QCD phase transition. Further checks for the models 
which incorporate the QGP phase transition are in progress. 

Acknowledgement: YGM is grateful to Prof. J.B. Natowitz for calling his 
attention to the $\Delta$-scaling in intermediate energy heavy ion collision. 

{}

\end{document}